# Refinement Checking for Multirate Hybrid ZIA

Guozheng Li[1], Zining Cao[1] and Zheng Gao[1]

[1]Computer Science Department

Nanjing University of Aeronautics & Astronautics, Nanjing, China

leebug38@foxmail.com

ABSTRACT. *A hybrid system is a dynamical system with both discrete and continuous components. In order to study the modeling and verification aspects of hybrid system, in this paper we first introduce a specification approach combining interface automata, initialized multirate hybrid automata and Z language, which is named MZIA. Meanwhile we propose a refinement relation on MZIAs. Then we give an algorithm for checking refinement relation between MZIAs with finite domain and demonstrate the correctness of the algorithm.*

1. **Introduction.** Modern software systems are comprised of numerous components, and are made larger through the use of software frameworks. Hybrid software/hardware systems exhibit various behavioral aspects such as discrete and continuous transition, communication between components, and state transformation inside components. To ensure the correctness of these processes, formal specification techniques for such systems have to be able to describe all these aspects. Unfortunately, a single specification technique that is well suited for all these aspects is yet not available. Instead one needs various specialized techniques that are very good at describing individual aspects of system behavior. This observation has led to research into the combination and semantic integration of specification techniques. In this paper we combine three well researched specification techniques: Interface automata, multirate hybrid automata and Z.

Interface automaton is a light-weight automata-based language for component specification, which was proposed in [1]. An interface automaton (IA), introduced by de Alfaro and Henzinger, is an automata-based model suitable for specifying component-based systems. Hybrid automaton [2] is a formal model for a mixed discrete-continuous system. Z [3] is a typed formal specification notation based on first order predicate logic and set theory.

In this paper, we introduce a specification language which combines interface automata, multirate hybrid automata and Z language, named MZIA. Roughly speaking, a MZIA is in a style of multirate hybrid interface automata but its states and operations are described by Z language. Then the refinement relation between MZIAs is defined. At last, we provide an algorithm for checking refinement relation on MZIAs with finite domain.

This paper is organized as follows: In next section, we propose a specification language-MZIA. The refinement relation for MZIA is presented in section 3. And we give the definition of MZIA with finite domain in section 4. In section 5, we give a refinement checking algorithm for MZIAs with finite domain. The paper is concluded in Section 6.

2. **Multirate Hybrid Interface Automata with Z.** In many cases, systems have both discrete and continuous property. To specify hybrid systems, we proposed the specification

ZIA and HZIA in [5,6]. In this paper, for the decidability of our refinement checking algorithm, we add some constrains to HZIA, named MZIA, which can be used to specify hybrid behavioral and the data structure aspects of a system as well.

**Definition 2.1.** *A multirate hybrid interface automata with Z (MZIA)* $P = \langle S_P, S_P^i, A_P^I,$ $A_P^O, A_P^H, X_P, V_P^I, V_P^O, V_P^H, C_P, F_P^S, F_P^A, I_P, T_P \rangle$ *consists of the following elements:*

*(1) $S_P$ is a set of states;*

*(2) $S_P^i \subseteq S_P$ is a set of initial states. If $S_P = \emptyset$ then P is called empty;*

*(3) $A_P^I$, $A_P^O$ and $A_P^H$ are disjoint sets of input, output, and internal actions, respectively. We denote by $A_P = A_P^I \cup A_P^O \cup A_P^H$ the set of all actions;*

*(4) $X_P = \{x_1, \ldots, x_n\}$ is a finite set of real-numbered variables; The number n is called the dimension of P. We write $\dot{X}_P = \{\dot{x}_1, \ldots, \dot{x}_n\}$ representing first derivatives during continuous change, and $X_P' = \{x_1', \ldots, x_n'\}$ representing values at the conclusion of discrete change. $\Re(X_P)$ is a rectangle over $X_P$;*

*(5) $V_P^I$, $V_P^O$ and $V_P^H$ are disjoint sets of input, output, and internal variables, respectively. We denote by $V_P = V_P^I \cup V_P^O \cup V_P^H$ the set of all variables. We have that $X_P \subseteq V_P$ which are all continuous valued variables and $V_P - X_P$ are all discrete valued variables;*

*(6) $C_P$ is a variable representing time, whose value is a real number, $C_P \notin V_P$;*

*(7) $F_P^S$ is a map, which maps any state in $S_P$ to a state schema $\Phi(V_P \cup \{C_P\})$ in Z language;*

*(8) $F_P^A$ is a map, which maps any input action in $A_P^I$ to an input operation schema $\Phi(V_P)$ in Z language, and maps any output action in $A_P^O$ to an output operation schema $\Phi(V_P)$ in Z language, and maps any internal action in $A_P^H$ to an internal operation schema $\Phi(V_P)$ in Z language;*

*(9) $I_P$ is a tuple $(inv_P, init_P, act_P)$, mapping from any state in $S_P$ to $\mathbb{Q}^n$ or $\Re(X_P)$, where $init_P : S_P \to \mathbb{Q}^n$ assigns an initial condition to each state, $inv_P : S_P \to \Re(X_P)$ assigns an invariant condition to each state, and $act_P : S_P \to \mathbb{Q}^n$ assigns a flow condition to each state $s \in S_P$ to indicate that $\dot{x} = act_P(s)$, for each $x \in X_P$;*

*(10) $T_P \subseteq S_P \times A_P \times \Re(X_P) \times 2^{X_P} \times \mathbb{Q}^n \times S_P$ is a set of transitions. The 6-tuple $(s, a, \varphi, \lambda, \xi, s') \in T_P$ corresponds to a transition from state $s$ to state $s'$ labeled with action $a \in A_P(s)$, a constraint $\varphi$ that specifies when the transition is enabled, and a set of real-numbered variables $\lambda \subseteq 2^{X_P}$ that are reset to the corresponding value in $\xi$ when the transition is executed. In this paper, we define that if for every coordinate $i \in \{1, \ldots, n\}$ with $act_P(s)_i \neq act_P(s')_i$, then $x_i \in \lambda$. Furthermore, we have $\models ((F_P^S(s) \wedge F_P^A(a)) \setminus (x_1, \cdots, x_m) \Leftrightarrow F_P^S(t)[y_1'/y_1, \cdots, y_n'/y_n])$, where $\{x_1, \ldots, x_m\}$ is the* set of variables in $F_P^S(s)$, $\{y_1,...,y_n\}$ is the set of the variables in $F_P^S(t)$, the set variables in $F_P^A(a)$ is the subset of $\{x_1,...,x_m\} \cup \{y_1,...,y_n\}$.

3. **Refinement Relation on MZIAs.** The refinement relation aims at formalizing the relation between abstract and concrete versions of the same component, for example, between an interface specification and its implementation. Roughly, a MZIA $P$ refines a MZIA $Q$ if all actions of $P$ can be simulated by $Q$.

In the following, we use $V^I(A)$ to denote the set of input variables in Z schema $A$, $V^O(A)$ to denote the set of output variables in Z schema $A$, $V^H(A)$ to denote the set of internal variables in Z schema $A$.

In order to define the refinement relation between Z schemas, we need the following notation.

**Definition 3.1.** *Consider two Z schemas $A$ and $B$ with $V^I(A) = V^I(B)$, $V^O(A) = V^O(B)$, $V^H(A) = V^H(B) = \emptyset$. We use the notation $A \geq B$ if one of the following cases holds:*

*(1) If $V^I(A) \neq \emptyset$ and $V^O(A) \neq \emptyset$ then given an assignment $\rho$ on $V^I(A)$, for any assignment $\sigma$ on $V^O(A)$, $\rho \cup \sigma \models B$ implies $\rho \cup \sigma \models A$, and given an assignment $\sigma$ on $V^O(A)$, for any assignment $\rho$ on $V^I(A)$, $\rho \cup \sigma \models A$ implies $\rho \cup \sigma \models B$;*

*(2) If $V^I(A) \neq \emptyset$ and $V^O(A) = \emptyset$ then for any assignment $\rho$ on $V^I(A)$, $\rho \models A$ implies $\rho \models B$;*

*(3) If $V^I(A) = \emptyset$ and $V^O(A) \neq \emptyset$ then for any assignment $\rho$ on $V^O(A)$, $\rho \models B$ implies $\rho \models A$;*

*(4) $V^I(A) = \emptyset$ and $V^O(A) = \emptyset$.*

Intuitively, $A \geq B$ means that schemas $A$ and $B$ have the same input variables and the same output variables, and schema $B$ has bigger domains of input variables but smaller ranges of output variables than schema $A$.

Now we give the refinement relation between Z schemas, which describe the refinement relation between data structures properties of states.

**Definition 3.2.** *Consider two Z schemas $A$ and $B$ we use the notation $A \unrhd B$ if*

*(1) $V^I(A) \subseteq V^I(B)$ and $V^O(A) \subseteq V^O(B)$;*

*(2) $A \setminus (x_1,...,x_m) \geq B \setminus (y_1,...,y_n)$, where $\{x_1,...,x_m\} = V(A) - V^I(A) - V^O(A)$ and $\{y_1,...,y_n\} = V(B) - V^I(A) - V^O(A)$.*

For example, $A \triangleq [x?: \mathbb{N}; y!: \mathbb{R} \mid x \text{ is a even number}; y! = \pi \times x?] \unrhd B \triangleq [x?: \mathbb{N}; u?: \mathbb{R}; y!: \mathbb{R}; v!: \mathbb{R}; z: \mathbb{N} \mid y! = 2\pi \times \lfloor x?/2 \rfloor; v! = z * u?]$.

In the following, we give a refinement relation between MZIAs. For MZIAs, a state has not only behavior properties but also data properties. Therefore this refinement relation involves both the refinement relation between behavior properties and the refinement relation between data properties.

**Definition 3.3.** *Given a MZIA $P$ and a state $(s, D_P) \in S_P \times \mathbb{R}$ at some point, where $\mathbb{R}$ the set of real numbers, transitions in $P$ is is as follows.*

(1) *delay-transitions:* $(s, D_P) \xrightarrow{d} (s, D_P + d)$, where $d \in \mathbb{R}^+$, provided that for every $0 \leq \Delta d \leq d$, the invariant $inv_P(s)$ holds for $D_P + \Delta d$.

(2) *action-transitions:* $(s, D_P) \xrightarrow{a} (s', D_P')$, where $a \in A_p$, provided that there is a transition $(s, a, \varphi, \lambda, \xi, s')$ such that $s$ satisfies $\varphi$ and a set of real-numbered variables $\lambda$ that are reset to the corresponding value in $\xi$.

**Definition 3.4.** *Consider two MZIAs P and Q. A binary relation* $\succeq_M \subseteq (S_P \times \mathbb{R}) \times (S_Q \times \mathbb{R})$ *is a simulation from Q to P, if for all states* $(s, D_P) \in S_P \times \mathbb{R}$*, there exists* $(t, D_Q) \in S_Q \times \mathbb{R}$ *such that* $(s, D_P) \succeq_M (t, D_Q)$ *the following conditions hold:*

(1) $F_P^S(s) \trianglerighteq F_Q^S(t)$;

(2) *For any delay* $d \in \mathbb{R}^+$, *if* $(s, D_P) \xrightarrow{d} (s, D_P + d)$, *there exists a delay* $d' \in \mathbb{R}^+$ *such that* $(t, D_Q) \xrightarrow{d'} (t, D_Q + d')$;

(3) *For any action* $a \in A_p$, *if* $(s, D_P) \xrightarrow{a} (s', D_P')$, *there exists a state* $(t', D_Q') \in S_Q \times \mathbb{R}$ *such that* $(t, D_Q) \xrightarrow{a} (t', D_Q')$ , $F_P^A(a) \trianglerighteq F_Q^A(a)$ , $F_P^S(s') \trianglerighteq F_Q^S(t')$ , *and* $(s', D_P') \succeq_M (t', D_Q')$.

**Definition 3.5.** *The MZIA Q refines the MZIA P written* $P \succeq_M Q$ *if there is a simulation* $\succeq_M$ *from Q to P, a state* $(s, 0) \in S_P^i$ *and a state* $(t, 0) \in S_Q^i$ *such that* $(s, 0) \succeq_M (t, 0)$.

4. **MZIA With Finite Domain.** Consider a MZIA $P$ and a pair $(s, D_P) \in S_P \times \mathbb{R}$, where $\mathbb{R}$ is the set of real numbers. Obviously, $P$ is an infinite state system. For the decidability of our refinement checking algorithm, we should first convert infinite-state to finite-state. To obtain a finite representation for infinite state space of MZIA, we give the definition of MZIA with finite domain in this section.

In [7], the author proposed a constraint system called multirate zone for the representation and manipulation of multirate hybrid automata state-spaces. A multirate zone is a conjunction of inequalities of the following types: $ax - by \prec c$, $x \prec c$, and $c \prec x$, where $\prec \in \{<, \leq\}$, $c \in \mathbb{Q}$. Furthermore, the author showed that a multirate zone can be represented by a difference constraint matrix (DCM) and also gave three operations on DCMs: intersection, variable reset, and elapsing of time and proved that DCMs keep closed to the three operations.

We use multirate zone as the basis for the infinite state-space exploring of multirate hybrid automata, as well as for MZIAs. Furthermore, we introduce DCM to realize the multirate zones in the computer expediently. Here we will introduce a class of MZIAs, for which refinement checking problem is decidable.

**Definition 4.1.** *Given a Z schema* $s \triangleq [v_1 : T_1; \ldots; v_m : T_m \mid P_1; \ldots; P_n]$, *we call it a Z schema with finite domain, if every discrete variable* $v_i$ *in any schema has finite possible value, i.e., each type* $T_i$ *has finite elements. Consider a MZIA* $P = \langle S_P, S_P^i, A_P^I, A_P^O, A_P^H, X_P, V_P^I, V_P^O, V_P^H, C_P, F_P^S, F_P^A, I_P, T_P \rangle$ *is called a MZIA with finite domain, if the following condition holds:*

*(1) for each* $s \in S_P$, $F_P^S(s)$ *is a Z schema with finite domain;*

*(2) for each* $a \in S_P$, $F_P^A(a)$ *is a Z schema with finite domain.*

As multirate hybrid automata can be represented by DCM, so we get the finite state-space of MZIA with finite domain easily.

5. **MZIA with Finite Domain.** In the previous section, we represent multirate zones by difference constraint matrix. So it is easy for us to realize the process of MZIA $P$ converting to MZIA with finite domain $\mathcal{F}(P)$. In this section we present an algorithm *RC* for checking refinement relation over MZIAs with finite domain.

Suppose $\mathcal{F}(P)$ and $\mathcal{F}(Q)$ are two MZIAs with finite domain, the algorithm is given as follows:

| | |
|---|---|
| $RC(P,Q) =$ <br>    for each $p \in S^i_{\mathcal{F}(P)}, q \in S^i_{\mathcal{F}(Q)}$ <br>    $R_{p,q} := RCS(p,q)$ <br>    return $(\vee_{p,q} R_{p,q})$ <br> $RCS(p,q) =$ <br>    $B_{p,q} := RCZ\left(F^S_{\mathcal{F}(P)}(p), F^S_{\mathcal{F}(Q)}(q)\right)$ <br>    if $\wedge_{a \in A(p,q)}(p \xrightarrow{a}\!\!\!\!\!\!/\;)$ then return $(B_{p,q})$ <br>    else $B := \wedge_{a \in A(p,q)} Match_a(p,q)$ <br>    return $(B \wedge B_{p,q})$ <br> $Match_{a \in A(p,q)}(p,q) =$ <br>    if $(p \xrightarrow{a})$ and $(q \xrightarrow{a}\!\!\!\!\!\!/\;)$ then return (false) <br>    $C_a := RCZ(F^A_{\mathcal{F}(P)}(a), F^A_{\mathcal{F}(Q)}(a))$ <br>    for each $(p \xrightarrow{a} p_i)$ and $(q \xrightarrow{a} q_j)$ <br>      $C_{i,j} := RCZ(F^S_{\mathcal{F}(P)}(p_i), F^S_{\mathcal{F}(Q)}(q_j))$ <br>      $D_{i,j} := RCS(p_i, q_j)$ <br>    return $(\wedge_i (\vee_j (C_a \wedge C_{i,j} \wedge D_{i,j})))$ <br> $RCZ(S,T) =$ <br>    if $(V^I(S) \not\subseteq V^I(T))$ or $(V^O(S) \not\subseteq V^O(T))$ then <br>      return (false) <br>    else <br>      $V_S := V(S) - V^I(S) - V^O(S)$ | $V_T := V(T) - V^I(S) - V^O(S)$ <br>    $E := RCL(S \setminus V_S, T \setminus V_T)$ <br>    return $(E)$ <br> $RCL(M,N) =$ <br>    if $(V^O(M) \ne V^O(N))$ then <br>      return (false) <br>    if $(V^I(M) = \emptyset)$ and $(V^O(M) = \emptyset)$ then <br>      return (true) <br>    if $(V^I(M) = \emptyset)$ and $(V^O(M) \ne \emptyset)$ then <br>      return $(TV(N \Rightarrow M))$ <br>    if $(V^I(M) \ne \emptyset)$ and $(V^O(M) = \emptyset)$ then <br>      return $(TV(M \Rightarrow N))$ <br>    if $(V^I(M) \ne \emptyset)$ and $(V^O(M) \ne \emptyset)$ then <br>      return $(TV(\forall v^I_1 : V^I_1; \ldots; v^I_m : V^I_m \cdot (N \Rightarrow M)$ <br>      $\wedge \forall v^O_1 : V^O_1; \ldots; v^O_n : V^O_n \cdot (M \Rightarrow N)$ <br>      where $V^I(M) = \{v^I_1, \ldots, v^I_m\}, V^O(M) =$ <br>      $\{v^O_1, \ldots, v^O_n\}$, the type of $v^I_k$ is $V^I_k$, and <br>      the type of $v^O_k$ is $V^O_k$. <br> $TV(LS) =$ <br>    *rewrite schema LS to an equivalent* <br>    *first order logical formula LF* <br>    *if (LF is always true for any assignment* <br>    *on variables) then return (true)* <br>    *else return (false)* |

Suppose that $p$ and $p'$ are states of $\mathcal{F}(P)$, and $a$ is an action of $\mathcal{F}(P)$, we use the notation $p \xrightarrow{a} p'$ to denote $(p, a, p') \in T_{\mathcal{F}(P)}$. $A(p) = \{a \mid \exists p' \cdot p \xrightarrow{a} p'$ and $a$ is an action$\}$. $A(p) \cup A(q)$ is abbreviated as $A(p,q)$. We use $p \xrightarrow{a}$ to represent there exists $p'$ such that $p \xrightarrow{a} p'$, where $a \in A(p,q)$. Since there are finite states in a MZIA, $p \xrightarrow{a}$ is decidable.

In the above algorithm, the function *TV(LS)* returns *true* if *LS* is always *true* for any assignment, otherwise returns *false*. In general, *TV(LS)* cannot be implemented since the tautology problem of first order logic is not decidable. But if the logic is restricted to some decidable sublogics, for example, each variable of a logical formula has finite possible values, the tautology problem of such logic becomes decidable. *TV(LS)* is decidable because *LS* is a Z schema with finite domain. There are only finite possible assignments on discrete variables and the constraints of continuous variables are rectangles.

**Lemma 5.1.** The algorithm *RCZ*(*S*,*T*) given in the above terminates and is correct, i.e., it returns true iff $S \trianglerighteq T$.

**Lemma 5.2.** The algorithm *RCS*(*p*,*q*) given in the above terminates and is correct, i.e., it returns true iff $p \succeq_M q$.

By the above Lemmas, we have the following proposition:

**Proposition 5.1.** The algorithm *RC*(*P*,*Q*) given in the above terminates and is correct, i.e., it returns true iff $P \succeq_M Q$.

6. **Conclusions.** In this paper, we define a combination of interface automata, multirate hybrid automata and Z called MZIA, which can be applied to specify the behavior and data structures properties of a hybrid system. In [5], we proposed the ZIA model, but this model cannot describe the hybrid properties. Then we proposed HZIA model in [6], but we didn't give the refinement checking algorithm. In [7], the author proposed the model checking procedure for the initialized multirate hybrid automata, but his model can't describe the behavior of the interfaces between components and the data structures properties of the system. Furthermore, we define the refinement relation for MZIA. At last, to verify systems, we provide a refinement checking algorithm for MZIAs with finite domain.

**Acknowledgment.** This paper was supported by the Aviation Science Fund of China under Grant No. 20128052064, the Fundamental Research Funds for the Central Universities under Grant No. NZ2013306 and National Basic Research Program of China (973 Program) under Grant No. 2014CB744903.

# Appendix
## 1. Proof of Lemma 5.1
**Proof.** By our definition of the data refinement relation, we have Lemma 5.1.

## 2. Proof of Lemma 5.2
**Proof.** The function *RCS(p,q)* starts with the initial pair (*p,q*), trying to check the similarity of *p* and *q* by matching transitions from them. While travelling the transition graph, at each pair of nodes the algorithm produces the outgoing transitions and next states according to the transition of MZIA. The transitions are then matched for simulation, and the algorithm goes on to the new state pairs if the matches are successful.

The function *Match$_a$*, performs a depth-first search on the product of the two labeled transition graphs. If one state fails to match another's transitions then they are not refinement and return *false*, otherwise return *true*.

The correctness of the algorithm for refinement relation is not difficult to justify. Each call of *Match$_a$(p,q)* performs a depth-first search in the product graph of the two transition graphs. This ensures that *Match$_a$(p,q)* can only be called for finitely many times since the states spaces of *P* and Q are finite. So we have Lemma 5.2.

## 3. Example
We demonstrate the procedure for the refinement checking with a simple example.
### 3.1 MZIA *P*
Boiler has been widely used in thermal power station, ships, industrial and mining enterprises and so on. Here we consider a simple boiler plant including temperature controller, boiler system and pressure monitor, as in Fig. 1. For convenience, we only consider the temperature and pressure in boiler system, and consider temperature monitor and pressure monitor as two components communicating with boiler system by some interfaces. To be specific, the boiler system will send the temperature value to the temperature monitor and the pressure value to the pressure monitor.

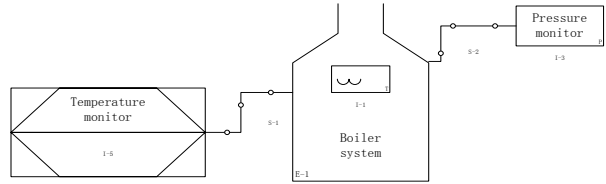

Fig. 1: Boiler plant.

Fig. 2 illustrates some states transitions of temperature and pressure, which is a MZIA model. The variable *x* and *y* represent the temperature and pressure, respectively. *x*! and *y*! mean the boiler system sends a output signal to the temperature monitor and the pressure monitor. Initially, we suppose that the temperature is 20(℃) and the pressure is standard atmospheric pressure, i.e. 100(Kpa). We omit the unit in the following. The automaton has four locations. The temperature and pressure are governed by derivatives in different location. The automaton starts in location $l_0$. It can remain in that location as long as the pressure is less than or equal to 1000. As soon as the pressure is greater than or equal to 700, the automaton can make a transition to location $l_1$ and reset the pressure to 700. Simultaneously, the derivative of the pressure is reset to 30. The rest of the transitions are similar.

We now see how the construction of the zones transitions described in the [7]. Multirate zones are represented by difference constraint matrix and the successor state is computed by the three operations on difference constraint matrix described above as well.

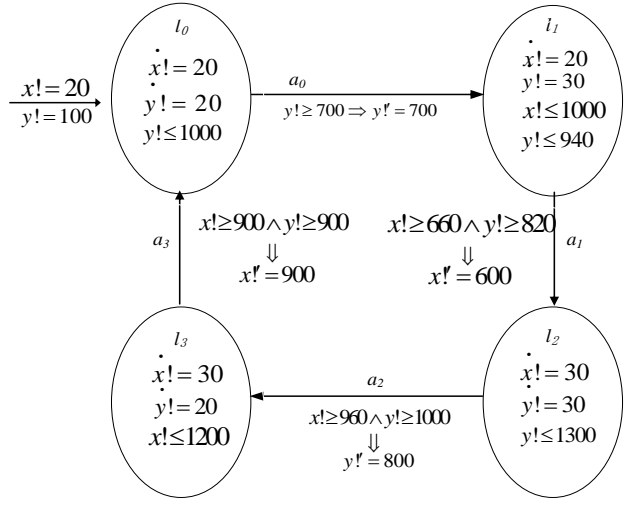

Fig. 2: States transitions of boiler system

Firstly, the initial state is given by $s_0 = (l_0, x? = 20 \land y! = 100)$ which corresponds to the difference constraint matrix $\mathcal{M}_0$:

|       | $x_0$            | $x!$              | $y!$              |
|-------|------------------|-------------------|-------------------|
| $x_0$ | $(1, 1, 0, \leq)$ | $(20, 1, -20, \leq)$ | $(20, 1, -100, \leq)$ |
| $x!$  | $(1, 20, 20, \leq)$ | $(1, 1, 0, \leq)$ | $(20, 20, -80, \leq)$ |
| $y!$  | $(1, 20, 100, \leq)$ | $(20, 20, 80, \leq)$ | $(1, 1, 0, \leq)$ |

We show the operation steps of getting next state with the intersection, variable reset, and elapsing of time operations. Only the canonical form DCM obtained in each step is shown.

(1) The invariant in location $l_0$ is $inv(l_0) = y! \leq 1000$, which is given by the matrix:

|       | $x_0$            | $x!$              | $y!$              |
|-------|------------------|-------------------|-------------------|
| $x_0$ | $(1, 1, 0, \leq)$ | $(20, 1, \infty, \leq)$ | $(20, 1, \infty, \leq)$ |
| $x!$  | $(1, 20, \infty, \leq)$ | $(1, 1, 0, \leq)$ | $(20, 20, \infty, \leq)$ |
| $y!$  | $(1, 20, 1000, \leq)$ | $(20, 20, \infty, \leq)$ | $(1, 1, 0, \leq)$ |

(2) Next, we let time elapse in the location l0 using the operator $\Uparrow$. The matrix for $(\mathcal{M}_0 \land inv(l_0))^{\Uparrow}$ is:

|       | $x_0$            | $x!$              | $y!$              |
|-------|------------------|-------------------|-------------------|
| $x_0$ | $(1, 1, 0, \leq)$ | $(20, 1, -20, \leq)$ | $(20, 1, -100, \leq)$ |
| $x!$  | $(1, 20, \infty, \leq)$ | $(1, 1, 0, \leq)$ | $(20, 20, -80, \leq)$ |
| $y!$  | $(1, 20, \infty, \leq)$ | $(20, 20, 80, \leq)$ | $(1, 1, 0, \leq)$ |

(3) The jump condition $\varphi = 700 \leq y!$ for the $a_0$ transition from location $l_0$ to location $l_1$ is:

|       | $x_0$            | $x!$              | $y!$              |
|-------|------------------|-------------------|-------------------|
| $x_0$ | $(1, 1, 0, \leq)$ | $(20, 1, \infty, \leq)$ | $(20, 1, -700, \leq)$ |
| $x!$  | $(1, 20, \infty, \leq)$ | $(1, 1, 0, \leq)$ | $(20, 20, \infty, \leq)$ |
| $y!$  | $(1, 20, \infty, \leq)$ | $(20, 20, \infty, \leq)$ | $(1, 1, 0, \leq)$ |

Furthermore, we intersect the set of states with the jump condition $\varphi$ to obtain $((\mathcal{M}_0 \wedge inv(l_0))^{\Uparrow} \wedge inv(l_0) \wedge \varphi)$:

|     | $x_0$ | $x!$ | $y!$ |
|-----|-------|------|------|
| $x_0$ | (1, 1, 0, $\leq$) | (20, 1, -620, $\leq$) | (20, 1, -700, $\leq$) |
| $x!$ | (1, 20, 920, $\leq$) | (1, 1, 0, $\leq$) | (20, 20, -80, $\leq$) |
| $y!$ | (1, 20, 1000, $\leq$) | (20, 20, 80, $\leq$) | (1, 1, 0, $\leq$) |

(4) Finally, we reset the variables in set $\lambda = \{y!\}$ to the corresponding value in set $\xi$. Here, $\xi_{y!} = 700$. So we obtain $\mathcal{M}_1 = [\lambda \mapsto \xi]((\mathcal{M}_0 \wedge inv(l_0))^{\Uparrow} \wedge inv(l_0) \wedge \varphi)$, which is given by the matrix:

|     | $x_0$ | $x!$ | $y!$ |
|-----|-------|------|------|
| $x_0$ | (1, 1, 0, $\leq$) | (20, 1, -620, $\leq$) | (30, 1, -700, $\leq$) |
| $x!$ | (1, 20, 920, $\leq$) | (1, 1, 0, $\leq$) | (30, 20, 13600, $\leq$) |
| $y!$ | (1, 30, 1000, $\leq$) | (20, 30, -4600, $\leq$) | (1, 1, 0, $\leq$) |

Note that the last difference constraint matrix corresponds to the multirate zone:
$s_1 = (l_1, 620 \leq x! \leq 920 \wedge 4600 \leq 30x! - 20y! \leq 13600 \wedge y! = 700)$ Consequently, the successor state in the mutirate zone automata is $s_1$. Repeating the same sequence of steps, we obtain the remaining states of the zone automata:

(1) $s_2 = (l_2, 820 \leq y! \leq 940 \wedge 6600 \leq 30y! - 30x! \leq 10200 \wedge x! = 600)$

(2) $s_3 = (l_3, 960 \leq x! \leq 1080 \wedge -4800 \leq 20x! - 30y! \leq -2400 \wedge y! = 800)$

(3) $s_4 = (l_0, 900 \leq y! \leq 960 \wedge 0 \leq 20y! - 20x! \leq 1200 \wedge x! = 900)$

(4) $s_5 = (l_1, 900 \leq x! \leq 1000 \wedge 13000 \leq 30x! - 20y! \leq 16000 \wedge y! = 700)$

(5) $s_6 = (l_2, 820 \leq y! \leq 850 \wedge 6600 \leq 30y! - 30x! \leq 7500 \wedge x! = 600)$

The reachability computation terminates at this point because the state $s_6$ is contained in $s_2$. Thus, no new states will be obtained by computing successor states in the zone automata.

Now we model the above boiler system based on our model MZIA with finite domain $P = \langle S_P, S_P^i, A_P^I, A_P^O, A_P^H, X_P, V_P^I, V_P^O, V_P^H, C_P, F_P^S, F_P^A, I_P, T_P \rangle$ which consists of the following elements:

(1) $S_P = \{s_0, s_1, s_1, s_2, s_3, s_4, s_5, s_6\}$;

(2) $S_P^i = \{s_0\}$;

(3) $A_P = \{a_0, a_1, a_2, a_3\}$;

(4) $V_P = \{x!; y!; l\}$;

(5) Here we introduce a global clock variable $clock \in C_P$ and map all states to their corresponding Z schema by function $F_P^S$. For the sake of space, we only use state $s_0$ and state $s_1$ as an example, and the rest states are similar: $F_P^S(s_0) \triangleq [l : \{l_0, l_1, l_2, l_3\}; x! : \mathbb{R}; y! : \mathbb{R}, clock : C_P | l = l_0; x! = 20; y! = 100; clock = 0]$; $F_P^S(s_1) \triangleq [l : \{l_0, l_1, l_2, l_3\}; x! : \mathbb{R}; y! : \mathbb{R}, clock : C_P | l = l_1; 620 \leq x! \leq 920; 4600 \leq 30x! - 20y! \leq 13600; y! = 700; 30 \leq clock \leq 45]$;

(6) Also, we map all actions to their corresponding Z schema by function $F_P^A$ with an example of action $a_0$: $F_P^S(a_0) \triangleq [y! : \mathbb{R} | y! = 700]$;

(7) $T_P = \{(s_0, a_0, s_1), (s_1, a_1, s_2), (s_2, a_2, s_3), (s_3, a_3, s_4), (s_4, a_0, s_5), (s_5, a_1, s_6)\}$.

### 3.2 **MZIA Q**

To illustrate the procedure for the refinement checking on MZIAs, we introduce another MZIA model in Fig. 3. Same as above, we get the states represented by DCMs as follows:

(1) $s'_0 = (l_0, x? = 20 \wedge y! = 100)$

(2) $s'_1 = (l_1, 720 \leq x! \leq 820 \wedge 7600 \leq 30x! - 20y! \leq 10600 \wedge y! = 700)$

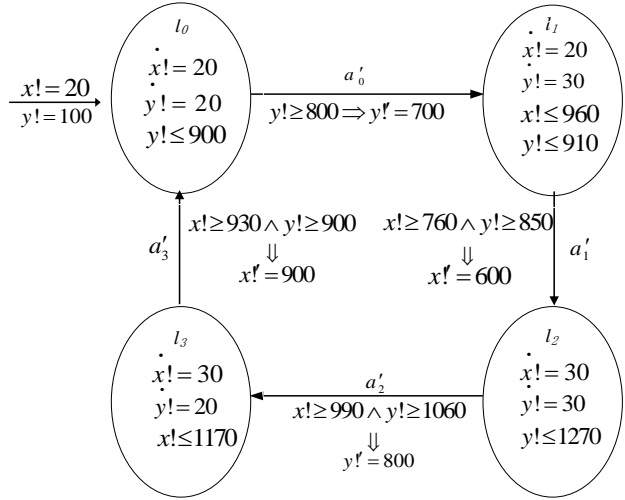

Fig. 3: States transitions of another boiler system

(3) $s'_2 = (l_2, 850 \leq y! \leq 910 \wedge 7500 \leq 30y! - 30x! \leq 9300 \wedge x! = 600)$

(4) $s'_3 = (l_3, 990 \leq x! \leq 1020 \wedge -4200 \leq 20x! - 30y! \leq -3600 \wedge y! = 800)$

(5) $s'_4 = (l_0, 900 \leq y! \leq 920 \wedge 0 \leq 20y! - 20x! \leq 400 \wedge x! = 900)$

Next we model the above system based on our MZIA with finite domain $Q = \langle S_Q, S_Q^i, A_Q^I, A_Q^O, A_Q^H, X_Q, V_Q^I, V_Q^O, V_Q^H, C_Q, F_Q^S, F_Q^A, I_Q, T_Q \rangle$ which consists of the following elements:

(1) $S_Q = \{s'_0, s'_1, s'_2, s'_3, s'_4\}$;

(2) $S_Q^i = \{s'_0\}$;

(3) $A_Q = \{a'_0, a'_1, a'_2, a'_3\}$;

(4) $V_Q = \{x!; y!; l\}$;

(5) We only use state $s'_0$ and state $s'_1$ as an example, and the rest states are similar:
$F_Q^S(s'_0) \triangleq [l : \{l_0, l_1, l_2, l_3\}; x! : \mathbb{R}; y! : \mathbb{R}, clock : C_Q \mid l = l_0; x! = 20; y! = 100; clock = 0]$; $F_Q^S(s'_1) \triangleq$
$[l : \{l_0, l_1, l_2, l_3\}; x! : \mathbb{R}; y! : \mathbb{R}, clock : C_Q \mid l = l_1; 720 \leq x! \leq 820; 7600 \leq 30x! - 20y! \leq 10600; y! = 700; 35 \leq clock \leq 40]$;

(6) Also, we map all actions to their corresponding Z schema by function $F_Q^A$ with an example of action $a_0$: $F_Q^S(a'_0) \triangleq [y! : \mathbb{R} \mid y! = 700]$;

(7) $T_Q = \{(s'_0, a'_0, s'_1), (s'_1, a'_1, s'_2), (s'_2, a'_2, s'_3), (s'_3, a'_3, s'_4)\}$.

### 3.3 **Refinement checking**

We illustrate the procedure for the refinement checking algorithm on MZIAs by checking $P \succeq_M Q$.

(1) For each $s_0 \in S_P^i$, $s_0' \in S_Q^i$, we have $F_P^S(s_0) \trianglerighteq F_Q^S(s_0')$, i.e. $RCZ\left(F_P^S(s_0), F_Q^S(s_0')\right)$ returns true.

(2) For $s_0 \xrightarrow{a_0} s_1$, we have $s_0' \xrightarrow{a_0'} s_1'$ such that $F_P^A(a_0) \trianglerighteq F_Q^A(a_0')$, $F_P^S(s_1) \trianglerighteq F_Q^S(s_1')$. Because $F_Q^S(a_0) \triangleq [y!: \mathbb{R} \mid y! = 700]$ and $F_Q^S(a_0') \triangleq [y!: \mathbb{R} \mid y! = 700]$, we have $F_P^A(a_0) \trianglerighteq F_Q^A(a_0')$. $F_P^S(s_1)$, $F_Q^S(s_1')$ have the same output variables $x!$ and $y!$, and schema $F_Q^S(s_1')$ has smaller ranges of output variables than schema $F_P^S(s_1)$.

(3) Repeat the above process, we get that $RC(P,Q)$ returns true, i.e., $P \succeq_M Q$.